\documentclass[showpacs,aps,prd,nofootinbib,superscriptaddress]{revtex4}
\usepackage{graphicx}
\usepackage{dcolumn}
\usepackage{amsmath}
\usepackage{epsfig}
\usepackage{colordvi}
\usepackage{color}
\usepackage{hhline}

\begin{document}
\title{\large \bf \boldmath
Study of $e^+e^- \to \omega \pi^0 \to
\pi^0\pi^0\gamma$  in the energy range 1.05--2.00 GeV with SND}

\author{M.~N.~Achasov}
\affiliation{Budker Institute of Nuclear Physics, SB RAS, Novosibirsk, 630090, Russia}
\affiliation{Novosibirsk State University, Novosibirsk, 630090, Russia}
\author{V.~M.~Aulchenko}
\author{A.~Yu.~Barnyakov}
\affiliation{Budker Institute of Nuclear Physics, SB RAS, Novosibirsk, 630090, Russia}
\author{K.~I.~Beloborodov}
\author{A.~V.~Berdyugin}
\affiliation{Budker Institute of Nuclear Physics, SB RAS, Novosibirsk, 630090, Russia}
\affiliation{Novosibirsk State University, Novosibirsk, 630090, Russia}
\author{A.~G.~Bogdanchikov},
\author{A.~A.~Botov}
\author{T.~V.~Dimova}
\affiliation{Budker Institute of Nuclear Physics, SB RAS, Novosibirsk, 630090, Russia}
\author{V.~P.~Druzhinin}
\author{V.~B.~Golubev}
\author{K.~A.~Grevtsov}
\author{L.~V.~Kardapoltsev}
\email[e-mail: ]{l.v.kardapoltsev@inp.nsk.su}
\author{A.~G.~Kharlamov}
\affiliation{Budker Institute of Nuclear Physics, SB RAS, Novosibirsk, 630090, Russia}
\affiliation{Novosibirsk State University, Novosibirsk, 630090, Russia}
\author{D.~P.~Kovrizhin}
\affiliation{Budker Institute of Nuclear Physics, SB RAS, Novosibirsk, 630090, Russia}
\author{I.~A.~Koop}
\author{A.~A.~Korol}
\affiliation{Budker Institute of Nuclear Physics, SB RAS, Novosibirsk, 630090, Russia}
\affiliation{Novosibirsk State University, Novosibirsk, 630090, Russia}
\author{S.~V.~Koshuba}
\affiliation{Budker Institute of Nuclear Physics, SB RAS, Novosibirsk, 630090, Russia}
\author{A.~P.~Lysenko}
\affiliation{Budker Institute of Nuclear Physics, SB RAS, Novosibirsk, 630090,
Russia}
\author{K.~A.~Martin}
\affiliation{Budker Institute of Nuclear Physics, SB RAS, Novosibirsk, 630090,
Russia}
\affiliation{Novosibirsk State Technical University,
Novosibirsk, 630092,Russia}
\author{I.~N.~Nesterenko}
\author{A.~E.~Obrazovsky}
\affiliation{Budker Institute of Nuclear Physics, SB RAS, Novosibirsk, 630090, Russia}
\author{E.~V.~Pakhtusova}
\affiliation{Budker Institute of Nuclear Physics, SB RAS, Novosibirsk, 630090, Russia}
\author{E.~A.~Perevedentsev}
\author{A.~L.~Romanov}
\author{S.~I.~Serednyakov}
\author{Z.~K.~Silagadze}
\affiliation{Budker Institute of Nuclear Physics, SB RAS, Novosibirsk, 630090, Russia}
\affiliation{Novosibirsk State University, Novosibirsk, 630090, Russia}
\author{K.~Yu.~Skovpen}
\author{A.~N.~Skrinsky}
\author{I.~K.~Surin}
\affiliation{Budker Institute of Nuclear Physics, SB RAS, Novosibirsk, 630090, Russia}
\author{Yu.~A.~Tikhonov}
\affiliation{Budker Institute of Nuclear Physics, SB RAS, Novosibirsk, 630090,
Russia}
\affiliation{Novosibirsk State University, Novosibirsk,
630090, Russia}
\author{A.~V.~Vasiljev}
\author{P.~Yu.~Shatunov}
\author{Yu.~M.~Shatunov}
\author{D.~A.~Shtol}
\affiliation{Budker Institute of Nuclear Physics, SB RAS, Novosibirsk, 630090,
Russia}
\author{A.~L.~Romanov}
\affiliation{Budker Institute of Nuclear Physics, SB RAS, Novosibirsk, 630090, Russia}
\affiliation{Novosibirsk State University, Novosibirsk, 630090, Russia}
\author{I.~M.~Zemlyansky}
\affiliation{Budker Institute of Nuclear Physics, SB RAS, Novosibirsk, 630090, Russia}

\begin{abstract}
The cross section for the process $e^+e^- \to \omega \pi^0 \to
\pi^0\pi^0\gamma$ has been measured in the center-of-mass energy 
range 1.05--2.00 GeV. The experiment has been performed at 
the $e^+e^-$ collider VEPP-2000 with the SND detector. The measured 
$e^+e^- \to \omega \pi^0$ cross section above 1.4 GeV is the most accurate
to date. Below 1.4 GeV our data are in good agreement with the previous SND
and CMD-2 measurements. Data on the $e^+e^-\to\omega\pi^0$ cross
section are well described by the Vector Meson Dominance~(VMD) model with 
two excited $\rho$-like states. From the measured cross section we have
extracted the $\gamma^*\to\omega\pi^0$ transition form factor. It
has been found that the VMD model cannot describe simultaneously
our data and data obtained from the $\omega\to\pi^0\mu^+\mu^-$
decay. We have also tested Conserved Vector Current~(CVC) hypothesis 
comparing our results on the $e^+e^- \to \omega \pi^0$ cross section 
with data on the $\tau^-\to\omega\pi^-\nu_{\tau}$ decay and have found
that the CVC hypothesis works well within reached experimental accuracy 
of about 5\%.
\end{abstract}

\maketitle

\section{Introduction}
Experiments with the SND detector~\cite{SND_desc} at the $e^+e^-$ 
collider VEPP-2000~\cite{vepp2k} started in 2010. The main goals of these
experiments are a high precision measurement
of the total cross section of $e^+e^-\to hadrons$
in the center-of-mass (c.m.) energy range up to 2 GeV and investigation 
of the vector meson excitations with masses between 1 and 2 GeV/$c^2$.
In this connection, a study of the process
\begin{equation}
\label{ompi} e^+e^- \to\omega \pi^0\to\pi^0\pi^0\gamma
\end{equation}
is very topical. The process
$e^+e^- \to \omega \pi^0$ is one of the dominant hadronic
processes contributing to the total hadronic cross section at the
c.m. energy between 1 and 2 GeV. As one of the important decay
modes of the isovector vector states $\rho(1450)$ and $\rho(1700)$, it can
provide a lot of information about their properties. Moreover,
this measurement can be used to check the relation between the
the $e^+e^- \to \omega \pi^0$ cross section and
the differential rate in the $\tau^-\to\omega\pi^-\nu_{\tau}$ decay
following from the conservation of vector current and isospin
symmetry (CVC hypothesis)~\cite{Tsai}. The SND plans to search for electric 
dipole decays
of the $\rho(1450)$ and $\rho(1700)$ mesons to the $\pi^0\pi^0\gamma$ final
state, which are important for
understanding of the $\rho(1450)$ and $\rho(1700)$ quark structure.
The process $e^+e^- \to \omega \pi^0$ is the main background for this
search; its precise measurement is the first step in investigating
the $\rho(1450)$ and $\rho(1700)$ radiative decays.

In this work the process $e^+e^- \to \omega \pi^0$ is studied
in the $\omega$ decay mode to $\pi^0\gamma$.
Despite the fact that the main $\omega$ decay mode
to $\pi^+\pi^-\pi^0$ has a probability
about an order of magnitude higher, this choice looks reasonable.
Unlike the $4\pi$ final state, the $\omega\pi^0$ intermediate mechanism is
dominant in the $\pi^0\pi^0\gamma$ final state in the energy
range under study. This makes it possible to avoid systematic
uncertainties due to both the complex procedure of subtracting
background and taking into account interference between different
intermediate mechanisms.

The process $e^+e^- \to \omega \pi^0$ in the $\omega\to\pi^0\gamma$
decay mode was first studied in the ND experiment~\cite{Dolinsky} at the 
VEPP-2M collider. The cross section was measured at c.m. energies below 1.4
GeV. Later this measurement was
repeated by the SND~\cite{ppg_SND} and CMD-2~\cite{ppg_CMD} detectors with much
higher statistics. In the energy region near the $\phi$-meson resonance
the cross section was measured in the SND experiment~\cite{ppg_SND_phi}
and then in the KLOE experiment~\cite{KLOE}.
Our measurement of the
$e^+e^- \to \omega \pi^0\to\pi^0\pi^0\gamma$ cross
section at VEPP-2000 based on 5 pb$^{-1}$ collected in 2010 was 
published in Ref.~\cite{SND2010}.

The first measurement of the process $e^+e^- \to \omega \pi^0$ in
the $\omega\to\pi^+\pi^-\pi^0$ decay mode was performed by the
DM2 Collaboration~\cite{dm2}. For a long time this measurement was the only 
one above 1.4 GeV. Below 1.4 GeV this cross section was measured at
VEPP-2M by SND~\cite{ppg_SND_phi,4pi_snd} and CMD-2~\cite{4pi_cmd}, and
in the KLOE experiment~\cite{KLOE} in the $\phi$-meson region.

\section{Experiment} \label{Exper}
   SND~\cite{SND_desc} is a general
purpose non-magnetic detector. Its main part is a spherical
three-layer NaI(Tl) calorimeter with 560 individual crystals per
layer and 90\% solid angle coverage. The calorimeter energy resolution for
photons is $\sigma_{E}/E_\gamma=4.2\%/\sqrt[4]{E_\gamma(GeV)}$, the angular
resolution $\simeq 1.5^\circ$. There is a tracking system around
the collider beam pipe based on a nine-layer drift chamber and
a one-layer proportional chamber with cathode-strip readout. Outside 
the calorimeter a muon detector consisting of proportional
tubes and scintillation counters is placed. An aerogel Cherenkov
counter located between the drift chamber and the calorimeter is
used for particle identification.

Experiments at VEPP-2000 started in 2010. During 2010--2012 the c.m.~energy
range $E=1.05$--2.00 GeV was scanned several times with a step of
20--25 MeV.
The total integrated luminosity collected by SND in this energy range is
about 40 pb$^{-1}$. This work is based on data (27 pb$^{-1}$)
recorded in 2010--2011.

During the experiment, beam energy  was determined
using measurements of the magnetic field in the collider
bending magnets. To fix the absolute energy scale,
scan of the $\phi(1020)$ resonance was performed and
its mass was measured. However, possible instability and
uncertainties in collider components may lead to
a sizable energy bias when the beam energy increases
from 0.5 to 1 GeV.
The uncertainty in the energy setting was investigated in 2012
in special runs, in which the beam energy was
measured using the Compton backscattering method~\cite{compton}.
Based on the results of these runs we conservatively estimate the
uncertainty in the c.m. energy to be 5~MeV.

\section{Luminosity measurement}
In this analysis the process
\begin{equation}
e^+e^- \to \gamma\gamma \label{gg}
\end{equation}
is used for luminosity measurement. Similar to the process under study
(\ref{ompi}), the normalizing process (\ref{gg}) does not contain charged
particles in the final state. The selection criteria for the process (\ref{gg})
are chosen in such a manner that some uncertainties on the cross section
measurement cancel as a result of the normalization.

For example, in the selection of five-photon events of the process under study
we require the absence of charged tracks in an event. This leads to loss of
signal events that contain beam-generated spurious tracks. The probability
of such a loss may reach several percent and strongly depends on experimental
conditions (vacuum pressure in the collider beam pipe,
beam currents, {\ldots}).
Since the same condition is used for selection of the normalization process,
the systematic uncertainty associated with beam-generated extra tracks
cancels. Moreover, events of both processes are selected by the same
hardware trigger. Therefore the uncertainty associated with the trigger
inefficiency cancels too.

To select events of the process (\ref{gg}), the following
selection criteria are used:
\begin{itemize}
\item at least two photons and no charged particles are detected,
\item the number of hits in the drift chamber is less than or equal to five,
\item the energies of two most energetic photons in an event
are larger than $0.3 E$,
\item the azimuth angles of these photons satisfy the
condition $||\phi_1-\phi_2| - 180^{\circ}| < 11.5^{\circ}$,
\item the polar angles of these photons satisfy the
conditions $|\theta_1+\theta_2 - 180^{\circ}| < 17.2^{\circ}$ and
$36^{\circ} < \theta_{1,2} < 144^{\circ}  $.
\end{itemize}
Photons are reconstructed as clusters in the electromagnetic
calorimeter with energies greater than 30 MeV not associated with 
charged tracks in the drift chamber.

The condition on the number of hits suppresses background from Bhabha events
with unreconstructed tracks in the drift chambers.
We do not expect any significant background from other $e^+e^-$ annihilation
processes in the energy region under study.
To estimate possible cosmic-ray background, data recorded during
7.5 hours in a special run without beams are analyzed.
No $e^+e^- \to \gamma\gamma$ candidates are selected.
We estimate that the fraction of cosmic events in the sample of
selected two-photon events is less than $2\times10^{-4}$ and conclude
that the cosmic-ray background is negligible.

To calculate the detection efficiency and the cross section
of the process (\ref{gg}), a Monte-Carlo (MC) event generator based
on Ref.~\cite{Berends} is used. The integrated luminosity measured for
each energy point is listed in Table~\ref{allres}. The theoretical uncertainty
on the cross section calculation is about 1\%. The systematic uncertainty on
the detection efficiency is estimated to be 2\%.

\section{Event Selection}

At the first stage of the analysis, five-photon events are selected with
following criteria:
\begin{itemize}
\item at least five photons and no charged particles are detected,
\item the number of hits in the drift chamber is less than or equal to five,
\item $E_{tot}/E > 0.5$, where $E_{tot}$ is the total energy
deposition in the calorimeter.
\end{itemize}

For events passing the preliminary selection, kinematic fits to
the $e^+e^-\to 5\gamma$ and $e^+e^-\to \pi^0\pi^0\gamma$
hypotheses are performed with requirements of energy and momentum
conservation and $\pi^0$ mass constraints for the second
hypothesis. The goodness of the fits is characterized by the
$\chi^2$ parameters: $\chi^2_{5\gamma}$ and
$\chi^2_{\pi^0\pi^0\gamma}$. For events with more than five
photons, all five-photon combinations are tested and the one with
minimal $\chi^2_{\pi^0\pi^0\gamma}$ is used. To select
$\omega\pi^0$ candidates the following additional conditions are
applied:
\begin{itemize}
\item $\chi^2_{5\gamma}<30$ for $E<1.7$ GeV and
$\chi^2_{5\gamma}<15$ for $E\ge 1.7$ GeV,
\item $\chi^2_{\pi^0\pi^0\gamma}-\chi^2_{5\gamma}<10$,
\item at least one of the two $\pi^0\gamma$ invariant masses
satisfies the condition $|m_{\pi^0\gamma}-M_{\omega}|<200$ MeV/$c^2$,
where $M_{\omega}$ is the $\omega$-meson nominal mass~\cite{pdg}.
\end{itemize}
\begin{figure}
\includegraphics[width=0.45\textwidth]{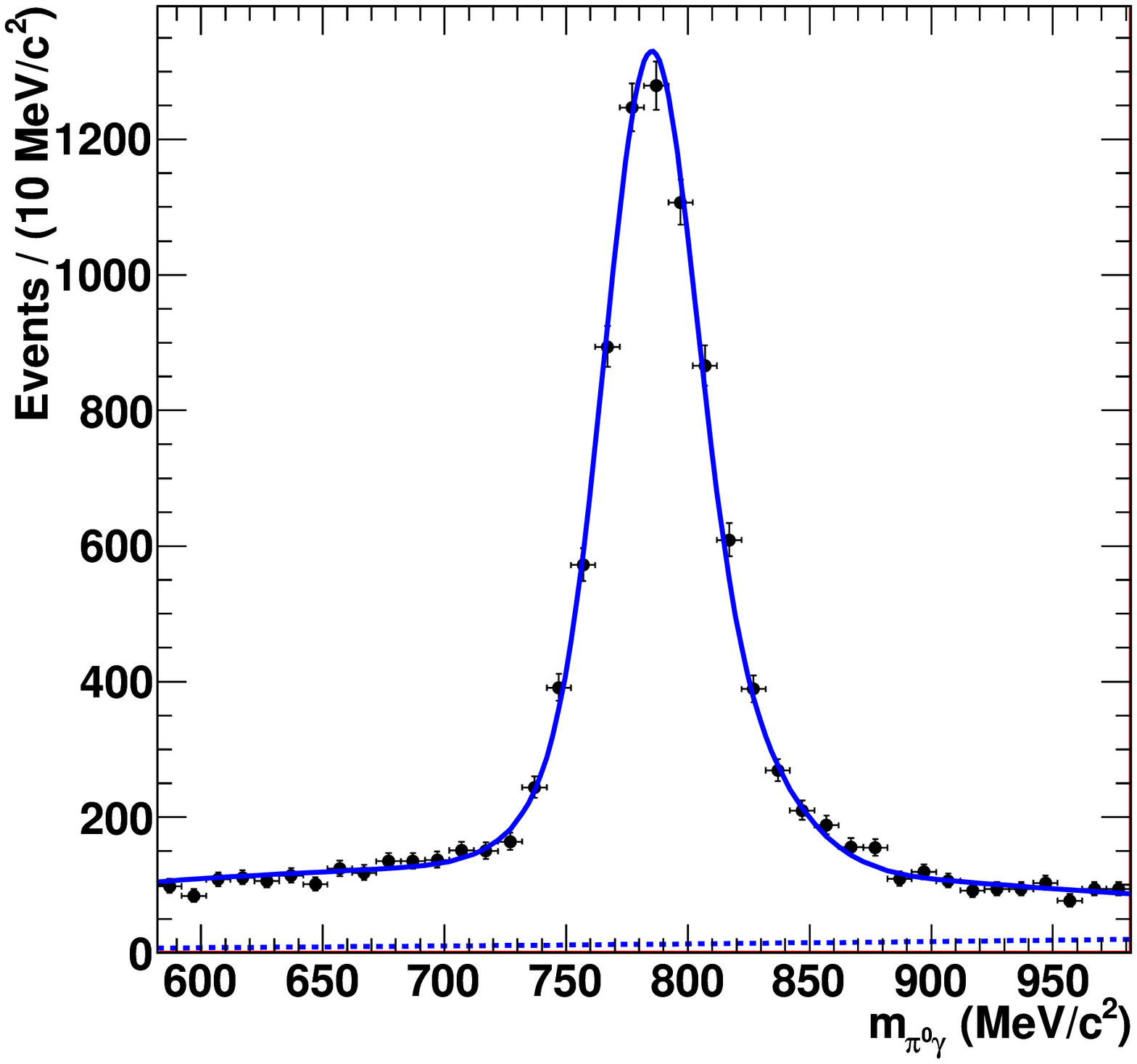} \hfill
\includegraphics[width=0.45\textwidth]{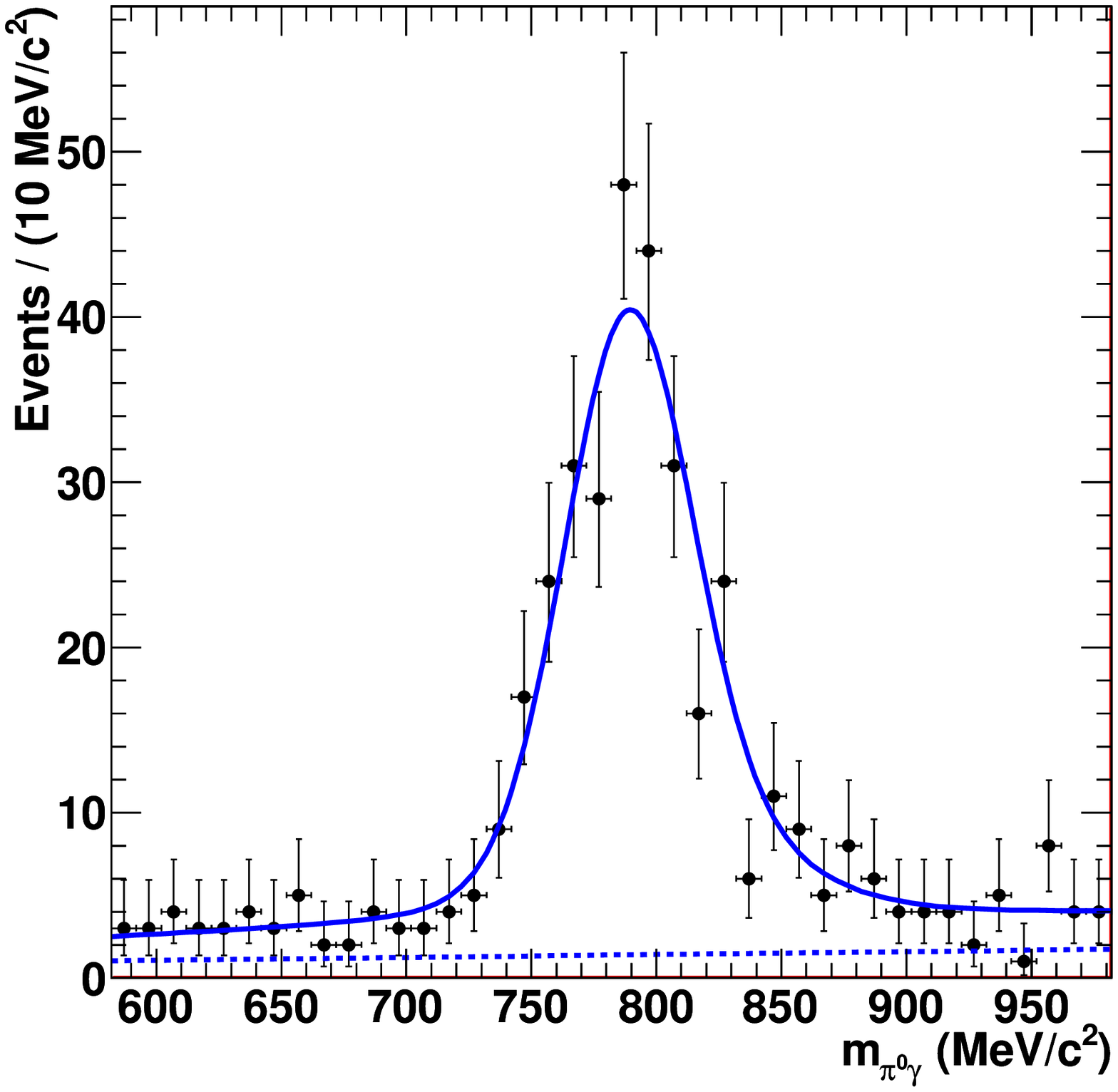} 
\caption{ The distribution of the $\pi^0\gamma$ invariant mass
for selected data events (points with error bars) with $E<1.7$ GeV (left)
and  $E\ge 1.7$ GeV (right).
The curves are the result of the fit described in the text.
The dashed line represents the linear-background contribution.\label{ompiM}}
\end{figure}

The distributions of the $\pi^0\gamma$ invariant mass for 7899 selected
data events at $E<1.7$ GeV and
331 data events at $E\ge 1.7$ GeV are shown in Fig.~\ref{ompiM}.
The $\omega$ meson peak is clearly seen in both distributions.
Since each event has two entries into the histogram, the nonresonant
parts of the distributions are determined mainly by signal events.

\section{Fitting the $\pi^0\gamma$ mass spectra}
To determine the number of signal events, the $\pi^0\gamma$ mass spectrum
is fitted by a sum of signal ($F_s$) and
background ($F_b$) distributions.
The signal distribution is obtained by fitting the mass spectrum for
simulated signal events with a non-parametric kernel estimation
technique~\cite{KEYS}.
To take into account a difference between data and MC simulation in mass
resolution and calibration, the $\pi^0\gamma$ mass for simulated events is
smeared and shifted before the fit. The values of the smearing Gaussian sigma
($\sigma_s$) and mass shift ($\Delta m$) are deduced from comparison of
the $\omega$ peak width and position in data with the same parameters in
simulation. To study energy dependence of $\sigma_s$ and $\Delta m$,
comparison is performed in three c.m.~energy regions below 1.7 GeV.
The mass shift and  the Gaussian sigma are found to be linearly dependent on the energy.
The value of $\Delta m$ changes from $(1.0 \pm 0.5)$ MeV/$c^2$ in the energy region
1.05--1.30 GeV to $(3.9\pm 0.5)$ MeV/$c^2$ in the energy region 1.5--1.7 GeV. The value of
$\sigma_s$ for the same energy regions changes from $(6.2 \pm 1.0)$ MeV/$c^2$
to $(10.5\pm 0.5)$ MeV/$c^2$.
For energies above 1.7 GeV, where statistics are small,
$\sigma_s$ is obtained by a linear extrapolation from lower energies.

The main sources of background are QED processes such as
$e^+e^-\to3\gamma$, $4\gamma$, $5\gamma$, and the hadronic
processes $e^+e^- \to \eta\gamma$ and $e^+e^- \to
\omega\pi^0\pi^0$. Background can arise also from the process
$e^+e^- \to \pi^0\pi^0\gamma$ with intermediate states other than
the $\omega\pi^0$, for example, $f_2\gamma$. The $\pi^0\gamma$
mass distribution for the QED, $\eta\gamma$ and $\pi^0\pi^0\gamma$
events is expected to be flat and is described by a linear
function. The distribution for $\omega\pi^0\pi^0$ events obtained
from MC simulation has a complex shape with a wide maximum shifted
to the right of the $\omega$ peak position. The expected
number of $\omega\pi^0\pi^0$ events is calculated using
experimental data on the $e^+e^- \to \omega\pi^+\pi^-$ cross
section~\cite{5pi} and the isotopic relation
$\sigma(\omega\pi^+\pi^-)=2\sigma(\omega\pi^0\pi^0)$. The
background from $e^+e^- \to \omega\pi^0\pi^0$ is important in the
energy range 1.7--1.9 GeV, but even there it does not exceed 6\% of
signal.

The cross section for the process $e^+e^-\to\rho^0\pi^0\to\pi^0\pi^0\gamma$ 
is estimated from the $e^+e^-\to\rho\pi\to3\pi$ cross section measured by 
BABAR~\cite{babar_3pi}, and found to be small, below 1 pb.
However, because of closeness of the $\rho$ and $\omega$ masses, 
the interference between the $\omega\pi^0$ and $\rho^0\pi^0$ amplitudes
can give a sizable contribution to the measured 
$e^+e^-\to\omega\pi^0\to\pi^0\pi^0\gamma$ cross section.
The $\pi^0\gamma$ spectrum for the interference term is 
peaked at $\omega$ mass and practically indistinguishable from the
spectrum for the $\omega\pi^0$ intermediate state. Since the phase between 
$\omega\pi^0$ and $\rho^0\pi^0$ amplitudes is unknown, we
calculate the maximum possible value of the interference term
and use this value as an estimate of the systematic uncertainty
on the measured $\omega\pi^0$ cross section. The uncertainty due to 
the interference is estimated to be 2\% at $E\le1.55$ GeV,
then increases to 4.5\% at 1.7 GeV, and
to 8.0\% at 1.8 GeV.  Above 1.8 GeV
this uncertainty is negligible in comparison with the uncertainty
of the radiative correction.

To determine the number of signal events we perform 
an unbinned extended likelihood fit to the mass spectrum in the range
$|m_{\pi^0\gamma}-M_{\omega}|<200$ MeV/$c^2$.
The likelihood function used for $E < 1.7$ GeV is given as follows:
\begin{multline}
L=P_P(N;N_s+N_b)P_B(M-N;N,k_t)\prod_{i=1}^{M}
\bigg(F_s(m_{\pi^0\gamma}^i)
\frac{N_s(1+k_s)}{N_s(1+k_s)+N_b(1+k_b)} +\\
F_b(m_{\pi^0\gamma}^i)
\frac{N_b(1+k_b)}{N_s(1+k_s)+N_b(1+k_b)}\bigg)
\end{multline}
where $N$ is the number of selected events,
$M$ is the total number of entries in the fitted spectrum,
$N_s$ ($N_b$) is the numbers of signal (background) events,
$k_s$ ($k_b$) is the fraction of signal (background) events with two entries
per event, i.e. events for which masses of both $\pi^0\gamma$ combinations
satisfy the condition $|m_{\pi^0\gamma}-M_{\omega}|<200$ MeV/$c^2$,
$k_t=(N_s k_s + N_b k_b)/(N_s+N_b)$.
The functions $P_P$ and $P_B$ are Poisson and binomial distributions for
the total number of selected events and the number of selected events with
two entries to the fitted spectrum, respectively.
The parameter $k_s$ is calculated using signal MC simulation.
It changes from 0.4 at $E=1.1$ GeV to 0.2 at $E=2.0$ GeV.
To understand the range of $k_b$ variation, we use MC simulation of
the $e^+e^-\to\pi^0\pi^0\gamma$ events at the generator level in
different models: according to phase space, with the intermediate
$f_0(980)\gamma$, $f_2(1270)\gamma$, and $f_0(1370)\gamma$ states.
In the energy range under study the $k_b$ value is found to change from
0.1 to 0.5. Therefore, the parameter $k_b$ is set to be $0.3\pm0.2$
in the fit (the likelihood function is multiplied by the corresponding Gaussian).
The background distribution
$F_b$ is described by a linear function. Above 1.7 GeV a term
describing $\omega\pi^0\pi^0$ contribution is added into
the likelihood function.

The fit results are shown in Fig.~\ref{ompiM}. To obtain the shape of
the mass spectrum for signal, the distribution of simulated events over energy
points is weighted to yield the distribution observed in data.
It is seen that our model for signal and background describes well
the experimental mass spectra. The total numbers of signal and background events
are $7533 \pm 110$ and $ 366 \pm 70$, respectively, for $E < 1.7$ GeV, and
$282 \pm 22$ and $49\pm 15$ for $E\ge 1.7$ GeV.
A similar fitting procedure is applied in each energy point. The numbers
of signal events obtained from the fits are listed in Table~\ref{allres}.

\section{Detection efficiency and radiative corrections}

\begin{figure}[]
\includegraphics[width=0.45\textwidth]{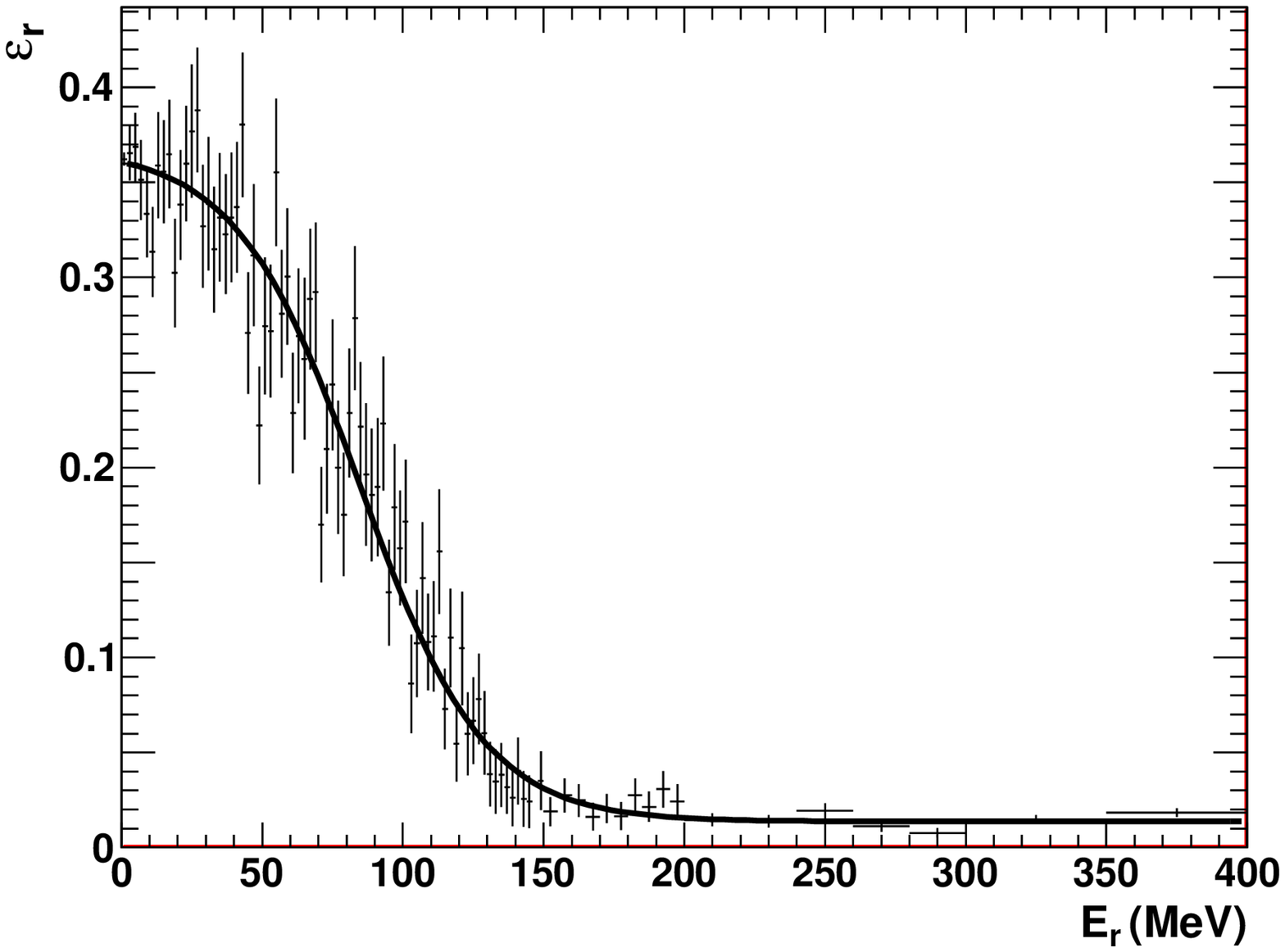} \hfill
\includegraphics[width=0.45\textwidth]{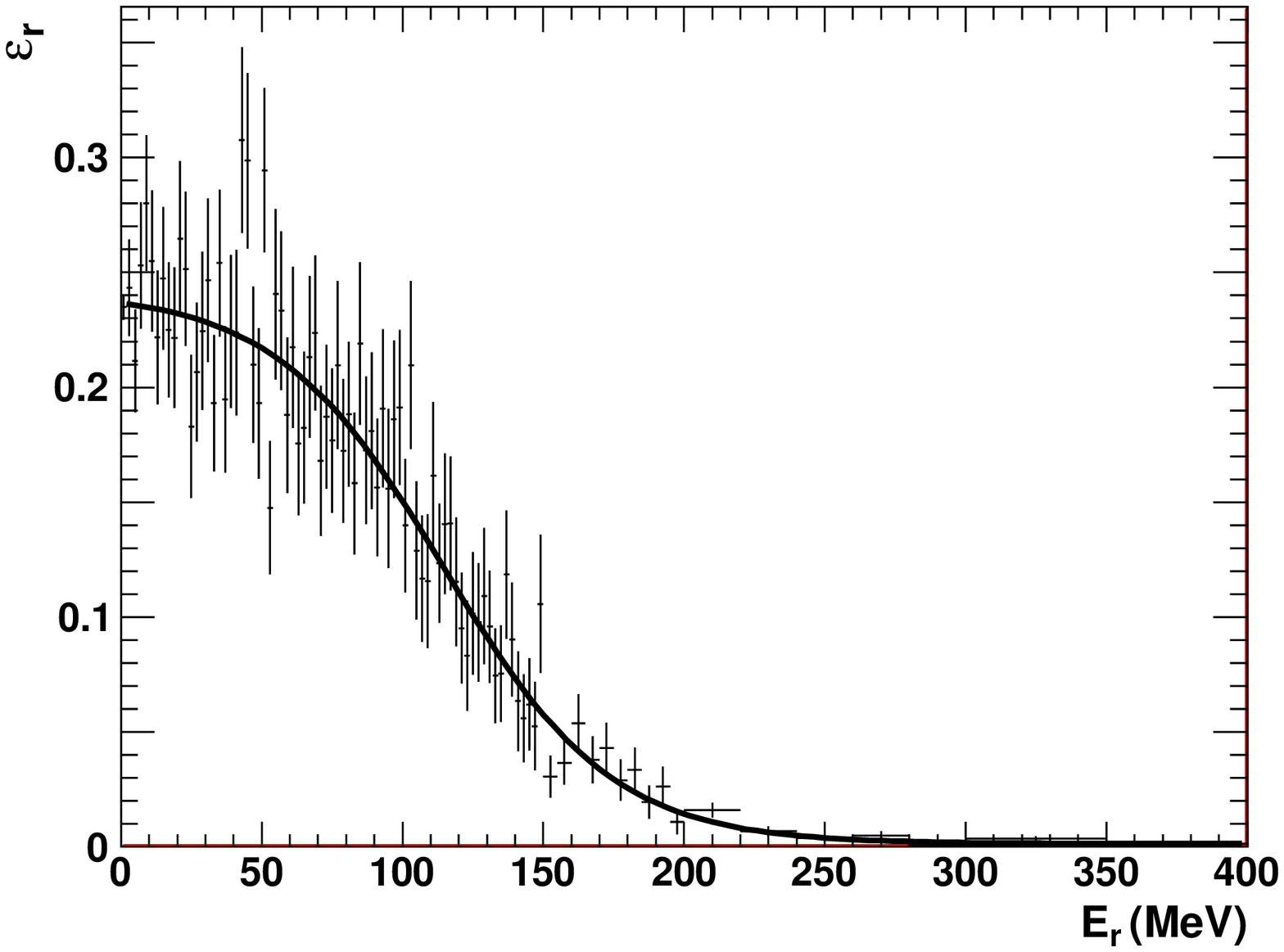}
\caption{The $E_r$ dependence of the detection efficiency for the process
$e^+e^- \to \omega \pi^0 \to \pi^0\pi^0\gamma$ for $E = 1.05$ GeV (left) and
$E = 2.00$ GeV (right).\label{EffRad}}
\end{figure}

The detection efficiency for the process under study is determined
using MC simulation. The simulation includes
radiative corrections to the Born cross section calculated
according to Ref.~\cite{FadinRad}. In particular, an extra photon
emitted by initial electrons is generated with the angular distribution
modeled according to Ref.~\cite{BoneMartine}. The
detection efficiency $\varepsilon_r$ is evaluated as a function of
two parameters: the c.m.~energy $E$ and the energy of the extra photon $E_r$.
The $E_r$ dependence of the detection efficiency is shown in
Fig.~\ref{EffRad} for the energy points with minimum and maximum energies
studied.

The visible cross section for the process $e^+e^- \to \omega \pi^0
\to \pi^0\pi^0\gamma$ is written as
\begin{equation}
\sigma_{vis} = \int \limits_{0}^{x_{max}}\varepsilon_r(E, xE/2)F(x,E)
\sigma(E\sqrt{1-x})dx,
 \label{viscrs}
\end{equation}
where $\sigma(E)$ is the Born cross section, which one needs to
extract from the experiment, $F(x,E)$ is a function describing the
probability to emit extra photons with the total energy $xE/2$~\cite{FadinRad}.
Equation~(\ref{viscrs}) can be rewritten in the conventional form:
\begin{equation}
\label{viscrs2} \sigma_{vis} = \varepsilon(E)\sigma(E)(1+\delta(E)),
\end{equation}
where $\delta(E)$ is the radiative correction, and $\varepsilon(E)$ is defined
as follows:
\begin{equation}
\label{EffDif} \varepsilon(E) \equiv \varepsilon_r(E,0).
\end{equation}
Technically the determination of the Born cross section is
performed as follows. With the use of Eq.~(\ref{viscrs}) the
energy dependence of the measured visible cross section is
approximated. To do this the Born cross section is parametrized by
some theoretical model that describes data reasonably well. The
fitted model parameters are used to evaluate the radiative correction
$\delta(E)$. Then the experimental values of the Born cross section are
obtained using Eq.(\ref{viscrs2}).  To estimate the model dependence
of the radiative correction, the model parameters
are varied in a wide range, with the condition that the approximation
quality remains acceptable.
The fit to the measured cross section is described in detail in
Sec.~\ref{finres}. The obtained values of the radiative correction are
listed in Table~\ref{allres}.
The systematic uncertainty associated with the radiative corrections
is estimated to be about 1\% at $E < 1.6$ GeV and increases up to 5\%
at $E= 1.7$ GeV. Above 1.7 GeV the radiative correction
becomes large and highly model dependent. For these energy points
we quote a range of $(1+\delta)$ variation.

\begin{figure}[]
\center
\includegraphics[width=0.48\textwidth]{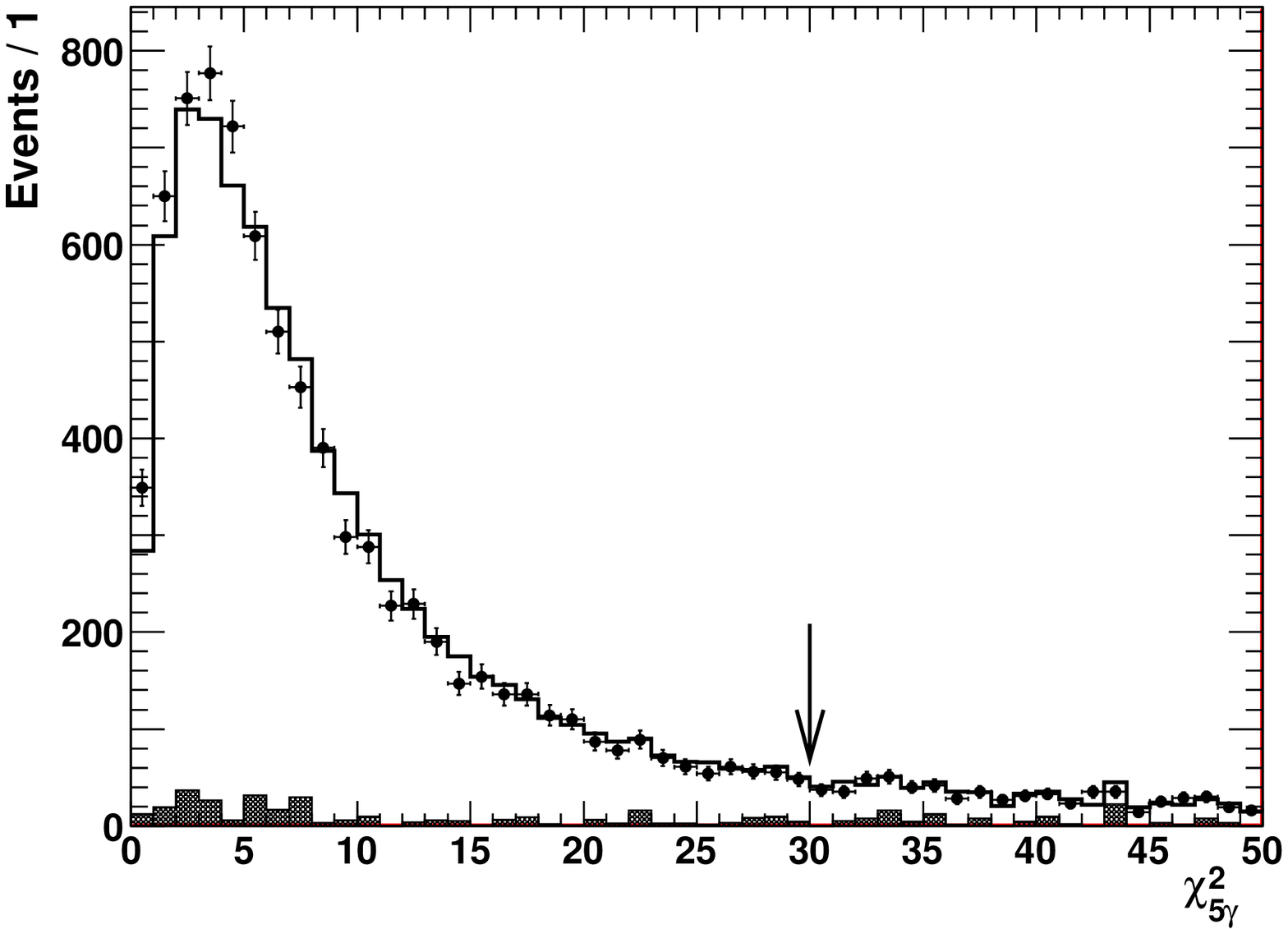} \hfill
\includegraphics[width=0.48\textwidth]{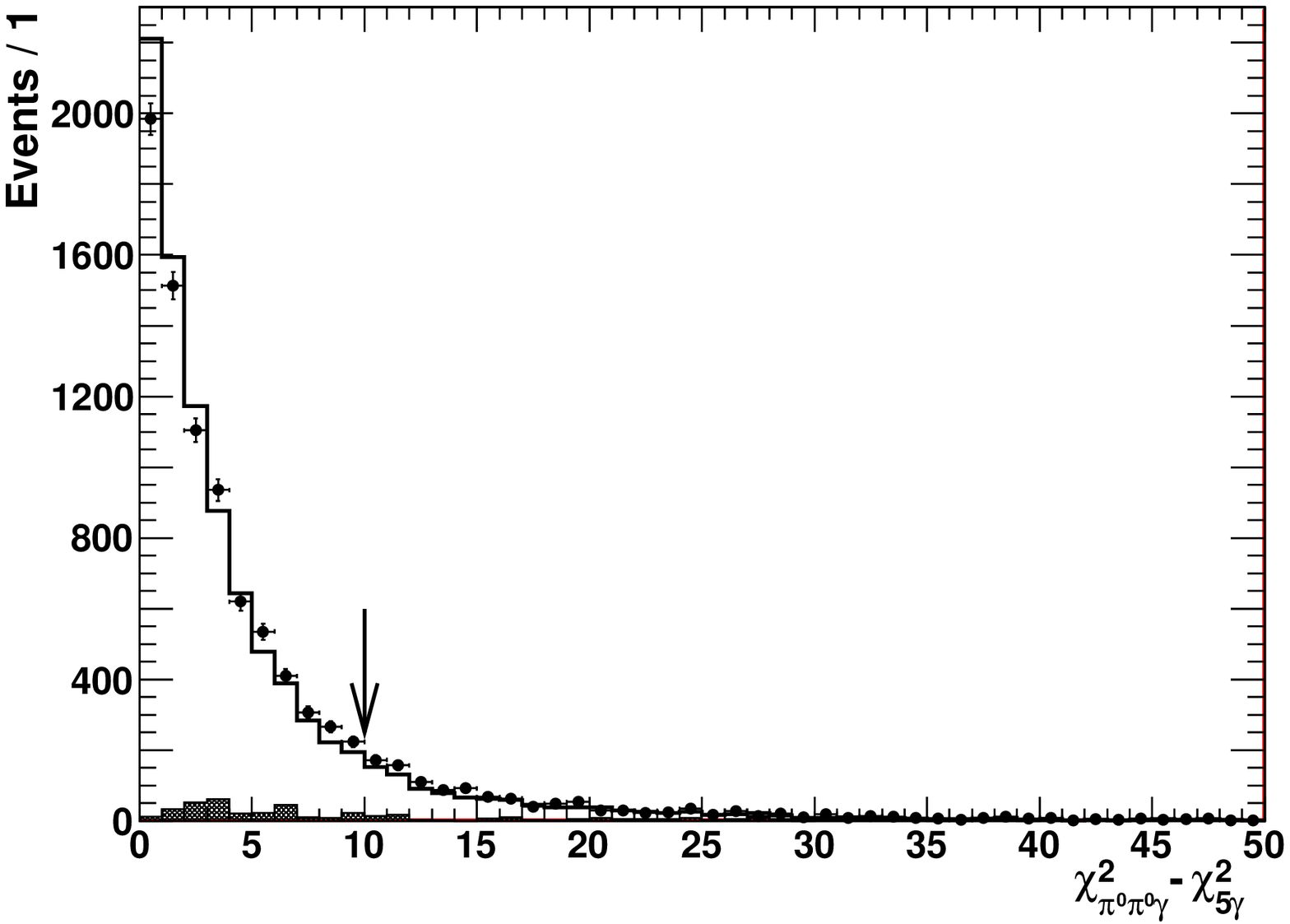}
\caption{The $\chi^2_{5\gamma}$ (left) and
$(\chi^2_{\pi^0\pi^0\gamma} - \chi^2_{5\gamma})$ (right) distributions
for data events with $E < 1.7$ GeV (points with error bars). The open
histogram is a sum of the simulated signal distribution and background
distribution. The latter is shown by the shaded histogram. The data and
the signal+background distributions are normalized to the same number of events.
The arrows indicate the selection criteria used.
\label{chi2}}
\end{figure}

%
Imperfect simulation of detector response for photons leads
to a systematic uncertainty in the detection efficiency.
To estimate this uncertainty, the data distributions of the
most important selection parameters $\chi^2_{5\gamma}$ and
$(\chi^2_{\pi^0\pi^0\gamma} - \chi^2_{5\gamma})$ for signal events
are compared with corresponding simulated distributions.
The distributions are shown in Fig.~\ref{chi2} for events
with $E < 1.7$ GeV. They are normalized to the same number of events.
The distributions for background events shown in Fig.~\ref{chi2} by the shaded histograms
are obtained using data from the $m_{\pi^{0}\gamma}$ sideband
($66.7\mbox{ MeV}/c^2 < |m_{\pi^{0}\gamma}-M_{\omega}| < 200.0\mbox{ MeV}/c^2$).
These background distributions are added to the simulated signal distributions.
A difference between data and MC simulation is seen in both the distributions. To obtain
numerical estimation, we change the limits of the conditions on $\chi^2_{5\gamma}$ and
$(\chi^2_{\pi^0\pi^0\gamma} - \chi^2_{5\gamma})$ from 30 to 50 and from 10 to 50,
respectively. The resulting variation of the measured cross section
$\delta\sigma_{\omega\pi}$ is found to be $(-0.4 \pm 0.5)\%$ for the $\chi^2_{5\gamma}$
condition, and $(2.9 \pm 0.5)\%$ for the $(\chi^2_{\pi^0\pi^0\gamma} - \chi^2_{5\gamma})$
condition. The obtained values of $\delta\sigma_{\omega\pi}$ are used to calculate
a correction to the detection efficiency. The detection efficiency obtained from
simulation for our standard selection criteria should be decreased by $(-2.5\pm0.7)\%$.

In the SND a photon converted in material before the drift chamber produces
a track. Events with converted photons
are rejected by the selection criteria used.
Since  the numbers of photons in the final state are different
for the signal and normalization processes,
the data-MC simulation difference in the conversion probability leads to
a systematic shift in the measured cross section. The conversion
probability is measured using $\phi \to \eta \gamma \to 3\gamma$ events
collected in a special run in vicinity of the $\phi$-meson resonance
and is found to be $(0.97 \pm 0.28)\%$ in data and $(0.78\pm0.04)\%$ in
simulation. The data-MC simulation difference $(0.19\pm 0.28)\%$
is consistent with zero. We
conclude that the simulation reproduces the photon conversion reasonably
well. The statistical error (0.3\%) of the data-MC simulation difference is
taken as an estimate of the systematic uncertainty on the photon conversion.
The corresponding uncertainty on the measured cross section is
estimated as $3\times0.3=0.9\%$ (the contributions from two of five photons
cancel due to normalization).

As it was discussed earlier, some part of the data events contains
beam-generated spurious tracks and photons. The effect of extra
tracks cancels due to normalization to two-photon events.
The presence of extra photons also changes the detection efficiency,
but differently for the signal and normalization processes.
To take into account this effect in MC simulation, beam-background events
recorded during experiment with a special random trigger are merged with
simulated events. The detection efficiencies obtained using simulation with
and without merged background are compared. It is found that the presence of
extra photons does not influence the number of selected normalization events.
The detection efficiency for the signal process increases by
0.3--1.3\% depending on experimental conditions. Unfortunately, the
random-trigger events were not recorded on a regular basis during the
2010--2011 experiments. Therefore, we conservatively estimate the correction
due to extra photons to the signal detection efficiency to be 0.8\% with
a systematic uncertainty of 0.5\%.

The corrected values of the detection efficiency $\varepsilon(E)$ are
listed in Table~\ref{allres}. The statistical error on the detection efficiency
is negligible. A systematic uncertainty from the sources discussed above
is 1.2\%. A nonmonotonic behavior of $\varepsilon(E)$
as a function of the c.m.~energy is due to variations of experimental
conditions, in particular, due to dead calorimeter
channels, a fraction of which changed from 2.4\% to 3.4\% during the data 
taking period.
\begin{table}
\caption{The c.m.~energy ($E$), integrated luminosity ($IL$), detection
efficiency ($\varepsilon$), number of selected signal events ($N_{s}$),
radiative-correction factor ($1+\delta$), measured Born cross section
($\sigma$). For the cross section the first error is statistical, the
second is systematic.\label{allres}}
\begin{ruledtabular}
\begin{tabular}{cccccc}
 $E$, GeV & $IL$, nb$^{-1}$ & $\varepsilon$, \% & $N_{s}$ & $1+\delta$ &
 $\sigma$, nb \\[0.3ex] \hline \\[-2.1ex]
1.050 &358 &35.5 & $104\pm 11$  & 0.903  &      $ 0.90\pm 0.10  \pm 0.03$     \\ 
1.075 &545 &36.5 & $176\pm 15$  & 0.913  &      $ 0.97\pm 0.08  \pm 0.03$     \\                     
1.100 &845 &36.0 & $297\pm 17$  & 0.921  &      $ 1.06\pm 0.06  \pm 0.04$     \\                     
1.125 &518 &35.9 & $220\pm 16$  & 0.928  &      $ 1.28\pm 0.09  \pm 0.04$     \\                     
1.150 &412 &37.8 & $178\pm 13$  & 0.934  &      $ 1.23\pm 0.09  \pm 0.04$     \\                     
1.175 &539 &36.8 & $231\pm 17$  & 0.939  &      $ 1.24\pm 0.09  \pm 0.04$     \\                     
1.200 &1058 &36.6 & $489\pm 24$  & 0.943  &     $ 1.34\pm 0.06  \pm 0.05$     \\                     
1.225 &550 &37.8 & $265\pm 19$  & 0.947  &      $ 1.35\pm 0.10  \pm 0.05$     \\                     
1.250 &435 &37.9 & $187\pm 17$  & 0.950  &      $ 1.19\pm 0.11  \pm 0.04$     \\                     
1.275 &495 &37.0 & $254\pm 21$  & 0.953  &      $ 1.46\pm 0.12  \pm 0.05$     \\                     
1.300 &1278 &37.6 & $673\pm 35$  & 0.956  &     $ 1.47\pm 0.08  \pm 0.05$     \\                     
1.325 &522 &38.2 & $279\pm 24$  & 0.959  &      $ 1.46\pm 0.12  \pm 0.05$     \\                     
1.350 &554 &38.1 & $292\pm 24$  & 0.962  &      $ 1.44\pm 0.12  \pm 0.05$   \\                       
1.375 &574 &38.1 & $302\pm 24$  & 0.966  &      $ 1.43\pm 0.11  \pm 0.05$     \\                     
1.400 &1012 &37.9 & $578\pm 33$  & 0.970  &     $ 1.55\pm 0.09  \pm 0.05$     \\                     
1.425 &598 &38.1 & $363\pm 26$  & 0.977  &      $ 1.63\pm 0.12  \pm 0.05$     \\                     
1.450 &427 &38.3 & $221\pm 19$  & 0.985  &      $ 1.37\pm 0.12  \pm 0.05$     \\                     
1.475 &599 &38.4 & $291\pm 21$  & 0.995  &      $ 1.27\pm 0.09  \pm 0.04$     \\                     
1.500 &1939 &39.0 & $996\pm 40$  & 1.007  &     $ 1.31\pm 0.05  \pm 0.04$     \\                     
1.525 &487 &38.2 & $245\pm 19$  & 1.021  &      $ 1.29\pm 0.10  \pm 0.05$     \\                     
1.550 &543 &38.2 & $228\pm 16$  & 1.038  &      $ 1.06\pm 0.08  \pm 0.04$     \\                     
1.575 &505 &37.9 & $170\pm 17$  & 1.063  &      $ 0.83\pm 0.08  \pm 0.04$ \\                         
1.600 &814 &38.1 & $232\pm 19$  & 1.100  &      $ 0.68\pm 0.06  \pm 0.03$     \\                     
1.625 &505 &37.9 & $139\pm 15$  & 1.161  &      $ 0.63\pm 0.07   ^{+0.03}_{-0.04 }$     \\           
1.650 &473 &36.9 & $96\pm 10$  &  1.262  &      $ 0.44\pm 0.04   ^{+0.02}_{-0.03 }$     \\           
1.675 &454 &37.0 & $74\pm 11$  &  1.429  &      $ 0.31\pm 0.05   ^{+0.01}_{-0.02 }$     \\           
1.700 &698 &30.3 & $70\pm 10$  &  1.704  &      $ 0.19\pm 0.05   ^{+0.01}_{-0.01 }$               \\           
1.725 &502 &30.6 & $22\pm 6$  & 2.0--2.3   &    $ 0.06\pm 0.04   ^{+0.007}_{-0.006 }$   \\           
1.750 &503 &29.2 & $25\pm 6$  & 2.4--3.3   &    $ 0.06\pm 0.04   ^{+0.02}_{-0.01 }$      \\          
1.775 &521 &28.7 & $22\pm 6$  & 2.8--5.0      & $ 0.03\pm 0.04   ^{+0.02}_{-0.01 }$       \\         
1.800 &727 &27.9 & $33\pm 7$  & 3.4--9.0      & $ 0.02\pm 0.04   ^{+0.02}_{-0.005 }$       \\        
1.825 &477 &28.2 & $7\pm 3$   & 4--15       & $0.004 ^{+0.027+0.008}_{-0.022-0.0 }$          \\       
1.850 &400 &26.7 & $4^{+4}_{-3}$ & 5--24    & $0.002 ^{+0.037+0.005}_{-0.027-0.0}$        \\          
1.870 &631 &26.2 & $19\pm 6$ & 5--27        & $0.005 ^{+0.038+0.016}_{-0.033-0.0}$          \\      
1.890 &577 &26.3 & $24\pm 5$ & 6--31        & $0.006 ^{+0.034+0.022}_{-0.031-0.001 }$       \\        
1.900 &553 &24.9 & $12^{+4}_{-5}$ & 6--40   & $0.004 ^{+0.028+0.011}_{-0.038-0.002 }$       \\     
1.925 &555 &24.6 & $14^{+4}_{-3}$ & 5--68   & $0.006 ^{+0.030+0.013}_{-0.025-0.005 }$       \\        
1.950 &406 &23.3 & $1^{+2}_{-1}$ & 5--63    & $0.001 ^{+0.021+0.001}_{-0.013-0.001}$        \\     
1.975 &460 &24.0 & $9^{+4}_{-5}$ & 5--39    & $0.009 ^{+0.033+0.008}_{-0.041-0.007 }$       \\        
2.000 &536 &23.3 & $5^{+3}_{-2}$ & 4--39    & $0.006 ^{+0.024+0.005}_{-0.018-0.005 }$       \\ \hline 
\end{tabular}
\end{ruledtabular}
\end{table}
\begin{figure}
\includegraphics[width=0.7\textwidth]{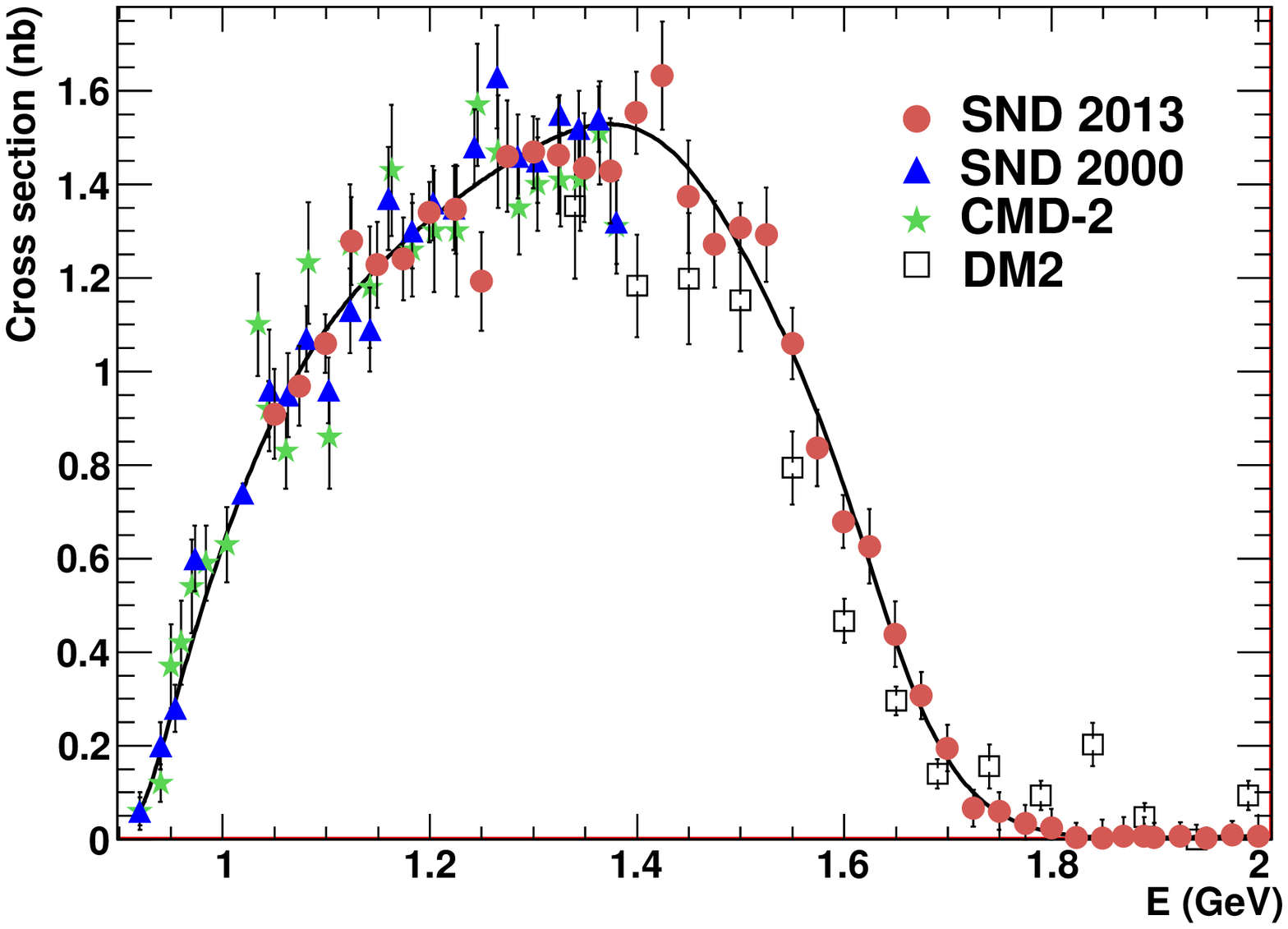}
\caption{The cross section for $e^+e^- \to \omega \pi^0 \to
\pi^0\pi^0\gamma$ measured in this work (circles), and in
SND~\cite{ppg_SND} (triangles),
CMD-2~\cite{ppg_CMD} (stars), and DM2~\cite{dm2} (squares) experiments.
Only statistical errors are shown.
The curve is the result of the fit to SND~2000 and SND~2013
data described in the text. \label{crs_ompi}}
\end{figure}

\section{Results and discussion\label{finres}}
The Born cross section for $e^+e^-\to \omega\pi^0\to \pi^0\pi^0\gamma$
obtained using Eq.(\ref{viscrs2}) is shown in Fig.~\ref{crs_ompi}
in comparison with the results of previous measurements. The numerical
values are listed in Table~\ref{allres}. The quoted errors on the cross
section are statistical and systematic. The sources of systematic
uncertainty are summarized in Table~\ref{syserr}.
The total systematic uncertainty is 3.4\% in the energy range $E\le1.55$ GeV and 
4.5\% in the energy range $1.55<E<1.6$.
Above 1.6 GeV the uncertainty increases due to the model dependence of
the radiative correction.

\begin{table}
\caption{The systematic uncertainties on the measured cross section
from different sources and the correction to the detection efficiency.
The total uncertainty is the sum of all the contributions in quadrature.
\label{syserr}}
\begin{ruledtabular}
\begin{tabular}{lcc}
 Source  & Systematic uncertainty, \% & Correction, \%  \\[0.3ex] \hline \\[-2.1ex]
Luminosity & 2.2  & ---    \\
Selection criteria  & 0.7 & -2.5\\
Photon conversion& 0.9 & ---\\
Beam background& 0.5 & 0.8 \\
Radiative correction ($E < 1.6$ GeV) & 1 & --- \\
Interference with $\rho^0\pi^0$ ($E < 1.6$ GeV) & 2-3.6 &
---\\
\hline Total & 3.4-4.5 & -1.7
\end{tabular}
\end{ruledtabular}
\end{table}

Our data are in good agreement with the measurements~\cite{ppg_SND,ppg_CMD} performed by SND and
CMD-2 at the VEPP-2M collider at energies below 1.4 GeV, but
significantly (by 20--30\%) exceed the DM2 data~\cite{dm2}. The DM2 data
were obtained in the $\pi^+\pi^-\pi^0\pi^0$ mode and have been rescaled using the
ratio of the $\omega\to\pi^0\gamma$ and $\omega\to 3\pi$ decay
probabilities~\cite{pdg}.

The cross section measured in this work is fitted together with the SND data
obtained in experiments at VEPP-2M~\cite{ppg_SND}. The Born cross section
is described by the following formula~\cite{ppg_SND}
\begin{equation}
\label{crsEq} \sigma(E)=
\frac{4\pi\alpha^2}{E^3}\left(\frac{g_{\rho\omega\pi}}{f_{\rho}}\right)^2
 \left|\frac{m_{\rho}^2}{D_{\rho}}+
A_1 e^{i\phi_1}\frac{m_{\rho^{\prime}}^2}{D_{\rho^{\prime}}}+ A_2
e^{i\phi_2}\frac{m_{\rho^{\prime\prime}}^2}{D_{\rho^{\prime
\prime}} }+
A_3\frac{m_{\rho^{\prime\prime\prime}}^2}{D_{\rho^{\prime
\prime\prime}} } \right|^2 P_f(E) B(\omega\to\pi^0\gamma),
\end{equation}
where $\alpha$ is the fine structure constant,
$g_{\rho\omega\pi}$ is the $\rho\to \omega\pi$ coupling constant,
$f_{\rho}$ is the $\gamma^{*}\to \rho$ coupling constant calculated from
the $\rho \to e^+e^-$ decay width,
$D_{\rho_i}(E) = m_{\rho_i}^2-E^2-\imath E \Gamma_{\rho_i}(E)$,
$m_{\rho_i}$ and $\Gamma_{\rho_i}(E)$ are the mass and width of the
resonance $\rho_i$, $B(\omega\to\pi^0\gamma)$ is the $\omega\to\pi^0\gamma$
branching fraction.
The first term in Eq.~(\ref{crsEq}) describes the contribution of
the $\rho(770)$ resonance, the second and third represent the
$\rho(1450)$ and $\rho(1700)$ contributions. The fourth term is
added to study a model dependence of the measured cross section
due to possible existence of a broad $\rho$-like resonance with mass
$m_{\rho^{\prime\prime\prime}} > 2$ GeV/$c^2$ or due to nonresonant contribution
into the Born cross section near 2 GeV.
The parameters $A_i$ are the ratios of the coupling constants
$A_i = g_{\rho_i\omega\pi}/g_{\rho\omega\pi}\cdot f_{\rho}/f_{\rho_i}$,
$\phi_1$ and $\phi_2$ are the phases of the $\rho(1450)$ and $\rho(1700)$
amplitudes relative to the $\rho(770)$ amplitude. For $\rho^{\prime\prime\prime}$
contribution, the phase is assumed to be equal to zero.
The energy dependence of the phase-space factor $P_f(E)$
is calculated using the MC event generator for signal events. For
an infinitely narrow $\omega$ resonance, $P_f(E) = 1/3\cdot
q_{\omega}^3$, where $q_{\omega}$ is the
$\omega$ meson momentum.
The energy dependence of the $\rho(770)$ width is described as follows:
\begin{equation}
\Gamma_{\rho}(E)=\Gamma_{\rho}(m_{\rho})
\left(\frac{m_{\rho}}{E}\right)^2
\left(\frac{q_{\pi}(E)}{q_{\pi}(m_{\rho})}\right)^3 +
\frac{g_{\rho\omega\pi}^2}{12\pi}q_{\omega}^3,
\end{equation}
where $q_{\pi}(E) = \sqrt{(E/2)^2-m_{\pi}^2}$,
$m_{\pi}$ is the $\pi^-$ mass. For $\rho(1450)$ and
$\rho(1700)$, the energy-independent widths are used.

The data are fit with free parameters $g_{\rho\omega\pi}$, $A_1$, $A_2$, $A_3$,
$M_{\rho^{\prime}}$, $ M_{\rho^{\prime\prime}}$, $\phi_1$,
and $\phi_2$. The values of the
$\rho^{\prime}$ and $\rho^{\prime\prime}$ widths are fixed at
PDG values~\cite{pdg}: $\Gamma_{\rho^{\prime}} = 400$ MeV,
$\Gamma_{\rho^{\prime\prime}} = 250$ MeV. The parameters
$M_{\rho^{\prime\prime\prime}}$ and $\Gamma_{\rho^{\prime\prime\prime}}$
are set to 2.3 GeV and 400 MeV, respectively.
The fit result is shown in Fig.~\ref{crs_ompi}.
The fitted parameters are used to calculate the values of the
radiative corrections listed in Table~\ref{allres}.
To estimate the model dependence of the radiative correction,
the fit is performed with $M_{\rho^{\prime}}$ fixed at
different values from the range 1.4--1.6 GeV and with
$A_3$ either free or fixed at zero.
The quality $P(\chi^2;\nu)$ of these fits, where $\nu$ is the number
of degrees of freedom, varies from 5 to 20\%.

To study the contributions of the $\rho(1450)$ and $\rho(1700)$ resonances,
we restrict the energy range to $E\le1.9$ GeV. This reduces the
model uncertainty due to a possible nonresonant contribution
or the $\rho^{\prime\prime\prime}$ resonance. The parameter
$A_3$ is set to zero.
The fit results are presented in Table~\ref{fitres}.
Two models have been studied, with a non-zero and zero
$\rho^{\prime\prime}$ contribution. In the fit with
the second model the parameter $\Gamma_{\rho^{\prime}}$ is left free.
Such a fit with only one $\rho$ excitation was performed in
Ref.~\cite{ppg_CMD} and gave a reasonable description of CMD-2 and DM2 data.
It is seen that our more precise data cannot be described by the model with one
excited $\rho$ state.
\begin{table}
\caption{Fit parameters obtained.\label{fitres}}
\begin{ruledtabular}
\begin{tabular}{ccc}
 Parameter & Model 1 & Model 2 \\ [0.3ex] \hline \\[-2.1ex]
 $g_{\rho\omega\pi}$, GeV$^{-1}$ & 15.6 $\pm$ 0.3 & 17.4 $\pm$ 0.1 \\
$A_1$ & 0.26 $\pm$ 0.01 & 0.11 $\pm$ 0.001 \\
 $A_2$ & 0.060 $\pm$ 0.006 & $\equiv$ 0\\
 $M_{\rho^{\prime}}$, MeV & 1491 $\pm$ 19 & 1488 $\pm$ 3 \\
$\Gamma_{\rho^{\prime}}$, MeV & $\equiv$ 400 & 321 $\pm$ 4 \\
$ M_{\rho^{\prime\prime}}$, MeV & 1708 $\pm$ 41 & --- \\
$ \Gamma_{\rho^{\prime\prime}}$, MeV & $\equiv$ 250 & --- \\
 $\phi_1$, deg. & 168 $\pm$ 3 & 121 $\pm$ 2\\
$\phi_2$, deg. & 10 $\pm$ 7& ---\\
 $\chi^2/\nu$ & 56.8 / 52& 118.6 / 54\\
\end{tabular}
\end{ruledtabular}
\end{table}
The fitted value of the parameter $A_i$ is used
to calculate the products of the branching fractions
\begin{equation}
B(\rho_i\to\omega\pi^0)\cdot B(\rho_i\to e^+e^-) =
\frac{\sigma_{\rho_i}(m_{\rho_i}) m^2_{\rho_i}}{12\pi},
\end{equation}
where
\begin{equation}
\sigma_{\rho_i}(E)=\frac{4\pi\alpha^2}{E^3}
\left(\frac{g_{\rho\omega\pi}}{f_{\rho}}\right)^2 \left|
A_i \frac{m_{\rho_i}^2}{D_{\rho_i}}\right|^2 P_f(E)
\end{equation}
is the cross section for the process $e^+e^-\to\rho_i\to\omega\pi^0$
without interference with other $\rho$-like resonances.
The results are following
\begin{equation*}
B(\rho^{\prime}\to e^+e^-)\cdot B(\rho^{\prime}\to\omega\pi^0) =
(5.3 \pm 0.4)\times 10^{-6},
\end{equation*}
\begin{equation}
\label{BB} B(\rho^{\prime\prime}\to e^+e^-)\cdot
B(\rho^{\prime\prime}\to\omega\pi^0) = (1.7 \pm 0.4)\times
10^{-6}.
\end{equation}
It should be noted that at the moment there is no generally accepted approach
for describing the tail of the $\rho(770)$ resonance above 1 GeV and shapes of broad
resonances like $\rho^{\prime}$ and $\rho^{\prime\prime}$.
The excitation curves of the three resonances $\rho$, $\rho^{\prime}$ and
$\rho^{\prime\prime}$ overlap; their amplitudes strongly
interfere with each other. As a result, a small change of the
resonance shape can lead to significant shifts in fitted resonance
parameters. So, the results obtained with our very simple model using
energy-independent $\rho^{\prime}$ and $\rho^{\prime\prime}$ widths
can be considered only as rough estimates of the resonance parameters.

The cross section for $e^+e^-\to \omega\pi^0$ can be expressed
in terms of the $\gamma^\ast\to \omega\pi^0$ transition form
factor $F_{\omega\pi\gamma}(q^2)$~\cite{lansberg,pacetti}, where $q$ is
the four-momentum of the virtual photon:
\begin{equation}
\sigma_{\omega\pi^0}(E)=\frac{4\pi\alpha^2}{E^3}|F_{\omega\pi\gamma}(E^2)|^2P_f(E).
\label{FormEq1}
\end{equation}
This form factor is also measured in the
$\omega\to\pi^0 e^+e^-$~\cite{CMDomega,SNDomega} and
$\omega\to\pi^0 \mu^+\mu^-$~\cite{LeptonG,NA60} decays. The value of
the form factor at $q^2=0$ is related to the
$\omega\to\pi^0\gamma$ partial width:
\begin{equation}
\Gamma(\omega\to\pi^0\gamma)=
\frac{\alpha}{3} P_\gamma^3| F_{\omega\pi\gamma}(0)|^2,
\label{FormEq2}
\end{equation}
where $P_\gamma$ is the decay photon momentum.
Using Eqs.~(\ref{crsEq}), (\ref{FormEq1}), (\ref{FormEq2}), and data from
Table~\ref{fitres} we calculate $\Gamma(\omega\to\pi^0\gamma)$ for
the model 1 to be $0.88\pm0.05$ MeV. For such a simple model the agreement
with the experimental value $\Gamma(\omega\to\pi^0\gamma)=0.703\pm0.024$
MeV~\cite{pdg} looks reasonable.

\begin{figure}
\includegraphics[width=0.7\textwidth]{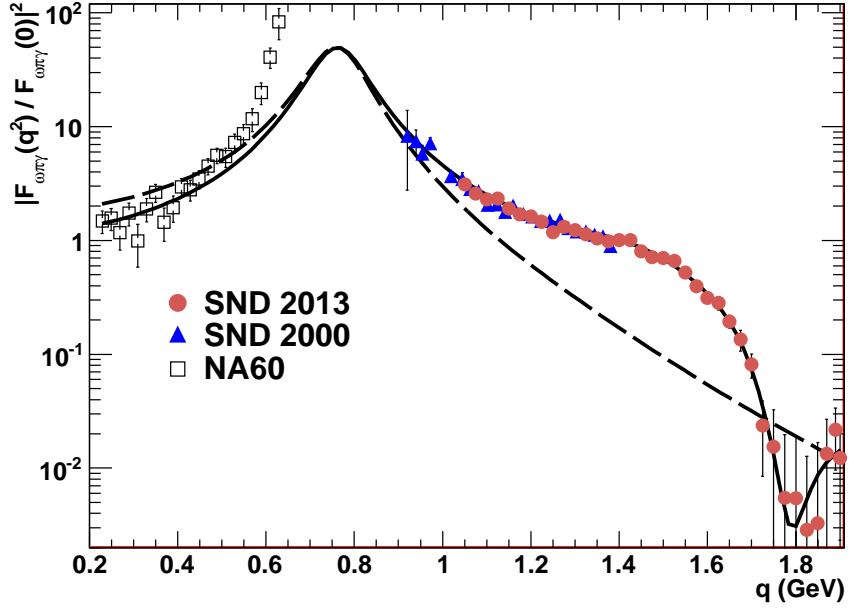}
\caption{The $\gamma^\ast\to\omega\pi^0$ transition form factor.
The points with error bars represent data from this work (circles),
Ref.~\cite{ppg_SND} (triangles), and Ref.~\cite{NA60} (squares).
Only statistical errors are shown.
The curve represent the result of model prediction with the parameters
listed in Table~\ref{fitres} for the model 1. The dashed curve shows the
$\rho(770)$ contribution. \label{form}}
\end{figure}

Figure~\ref{form} shows the normalized transition form factor squared
($|F_{\omega\pi\gamma}(q^2)/F_{\omega\pi\gamma}(0)|^2$) measured in this
work and in Ref.~\cite{ppg_SND} together with most precise data from omega
decays obtained in the NA60 experiment~\cite{NA60}. The curve represents the
results of the model prediction with the parameters listed in Table~\ref{fitres}
for model 1. The dashed curve shows the $\rho(770)$ contribution.
We conclude that it is hard to describe
data from $e^+e^-$ annihilation and the $\omega\to\pi^0\mu^+\mu^-$ decay
simultaneously with our model based on vector meson dominance (VMD).

\begin{figure}
\includegraphics[width=0.7\textwidth]{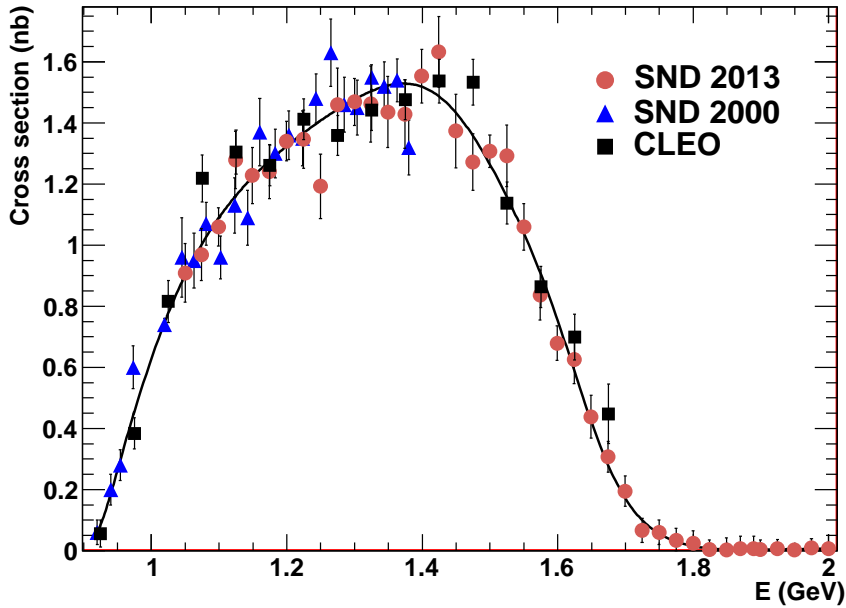}
\caption{ The cross section for $e^+e^- \to \omega
\pi^0 \to \pi^0\pi^0\gamma$ measured in this work (circles) and
in Ref.~\cite{ppg_SND} (triangles).
Only statistical errors are shown.
The cross-section data shown
by squares was calculated under CVC hypothesis from the spectral
function of the
$\tau \to \omega \pi \nu_{\tau}$ decay measured in CLEO
experiment~\cite{cleo}. The curve is the result of the fit to
SND~2013 and SND~2000 data. \label{tau}}
\end{figure}

The conserved vector current (CVC) hypothesis establishes a
relation between the charged hadronic current in the $\tau$ decay
and the isovector part of the electromagnetic current. So,
the $e^+e^-\to\omega\pi^0$ cross section can be related with
the spectral function of the $\tau^-\to\omega\pi^-\nu_{\tau}$ decay
($V_{\omega\pi}$)~\cite{Tsai}:
\begin{equation}
\sigma_{\omega\pi^0}(E)=\frac{4\pi^2\alpha^2}{E^2}V_{\omega\pi}(E).
\end{equation}
The comparison of the $e^+e^-\to\omega\pi^0\to\pi^0\pi^0\gamma$
cross section measured by SND with the cross section calculated under the
CVC hypothesis
from the $\tau^- \to \omega \pi^- \nu_{\tau}$ spectral function
measured in the CLEO experiment~\cite{cleo} is presented in Fig.~\ref{tau}.
It is seen that the $e^+e^-$ and $\tau$ data are in reasonable
agreement. The $\chi^2/\nu$ ($\nu$ is the number of degrees of freedom)
of comparison between the CLEO data and our fitted
curve is $19.7/16$. To calculate this $\chi^2$, 
a 5\% systematic uncertainty of CLEO data~\cite{cleo} was taken into account.

A more quantitative test of the CVC hypothesis can be made
by comparing the measured $\tau \to \omega \pi \nu_{\tau}$ branching
fraction with the value calculated from the $e^+e^-\to\omega \pi^0$
cross section according to the formula~\cite{Tsai,cleo}
\begin{equation}
\label{tauBr}
\Gamma(\tau^-\to\omega\pi^-\nu_{\tau})=
\frac{G_F^2 |V_{ud}|^2}{64\pi^4 \alpha^2
m_{\tau}^3}\int^{m_{\tau}}q^3(m_{\tau}^2-q^2)^2(m_{\tau}^2+2q^2)\sigma_{\omega\pi^0}(q)dq,
\end{equation}
where $|V_{ud}|$ is the Cabibbo-Kobayashi-Maskawa matrix element,
$m_{\tau}$ is the $\tau$ lepton mass, $G_F$ is the Fermi constant.
We integrate the fitted curve shown in Fig.~\ref{crs_ompi} and
obtain the value of the product
$\Gamma(\tau^-\to\omega\pi^-\nu_{\tau})B(\omega\to\pi^0\gamma)=
(3.68\pm0.04\pm0.13)\times 10^{-6}$ eV.
Using the values of the $\tau$ lifetime and
$B(\omega\to\pi^0\gamma)$ we calculate the branching fraction
\begin{equation}
\label{tauBr2} B(\tau^-\to\omega\pi^-\nu_{\tau}) = (1.96 \pm 0.02 \pm 0.10)
\times 10^{-2}.
\end{equation}
which is in good agreement with the experimental value
$(1.95 \pm 0.08) \cdot 10^{-2}$ obtained as a difference of
the PDG~\cite{pdg} values for $B(\tau^-\to\omega h^-\nu_{\tau})$
and $B(\tau^-\to\omega K^-\nu_{\tau})$.
We conclude that the CVC hypothesis for the $\omega\pi$ system
works well within the reached experimental accuracy of about 5\%.

\section{Summary}
The cross section for the
$e^+e^-\to\omega\pi^0\to\pi^0\pi^0\gamma$ process has been
measured with the SND detector at the VEPP-2000 $e^+e^-$
collider in the energy range of 1.05--2.00 GeV. This is the most accurate
measurement of the $e^+e^-\to\omega\pi^0$ cross section between 1.4 and
2.0 GeV. Below 1.4 GeV our data agree with the earlier measurements of the
same reaction performed at the VEPP-2M collider with the SND~\cite{ppg_SND}
and CMD-2~\cite{ppg_CMD} detectors. Significant disagreement is observed
with DM2 data~\cite{dm2} in the energy range 1.3--2.0 GeV.

Data on the $e^+e^-\to\omega\pi^0$ cross section are well described by
the VMD model with the three $\rho$-like state: $\rho(770)$, $\rho^{\prime}$
and $\rho^{\prime\prime}$.
However, the full data set on the $\gamma^\ast\to\omega\pi^0$ transition form
factor including data from both  $e^+e^-$ annihilation and $\omega$
decays, in particular, $\omega\to\pi^0\mu^+\mu^-$~\cite{NA60}, cannot be
described by such a simple model.

We have also tested the CVC hypothesis comparing the energy dependence of
the $e^+e^-\to\omega\pi^0$ cross section with the spectral function for the
$\tau^-\to\omega\pi^-\nu_{\tau}$ decay, and calculating the branching fraction
for this decay from $e^+e^-$ data. We have concluded that the CVC hypothesis
for the $\omega\pi$ system works well within the experimental accuracy of 5\%.

We would like to thank S.Ivashyn for useful discussions. This work
is supported by the Ministry of Education and Science of the
Russian Federation (Contract 14.518.11.7003), the Russian Federation
Presidential Grant for
Scientific Schools NSh-5320.2012.2, RFBR (grants 11-02-00276-a, 12-02-00065-a, 13-02-00418-a, 13-02-00375-a
12-02-31488-mol-a, 12-02-31692-mol-a, 12-02-31488-mol-a, 12-02-33140-mol-a-ved),the
Russian Federation Presidential Grant for Young Scientists
MK-4345.2012.2 and the Grant 14.740.11.1167 from the Federal Program ``Scientific and 
Pedagogical Personnel of Innovational Russia''.


\begin{thebibliography}{99}

\bibitem{vepp2k}
Yu.~M.~Shatunov {\it et al.,} in Proceedings of the 7th European
Particle Accelerator Conference, Vienna, 2000, p.439.
\bibitem{SND_desc}
M.~N.~Achasov {\it et al.,} Nucl. Instrum. Meth. A {\bf 598}, 31 (2009);
V.~M.~Aulchenko {\it et al.,} {\it ibid.} A {\bf 598}, 102 (2009);
A.~Yu.~Barnyakov {\it et al.,} {\it ibid.} A {\bf 598}, 163 (2009);
V.~M.~Aulchenko {\it et al.,} {\it ibid.} A {\bf 598}, 340 (2009).
\bibitem{Tsai}Y.-S.~Tsai,
Phys.\ Rev.\ D {\bf 4}, 2821 (1971)
[Erratum-ibid.\ D {\bf 13}, 771 (1976)].
\bibitem{Dolinsky}S.~I.~Dolinsky {\it et al.,}
Phys.\ Lett.\ B {\bf 174}, 453 (1986).
\bibitem{ppg_SND}M.~N.~Achasov {\it et al.} (SND Collaboration),
Phys.\ Lett.\  B {\bf 486}, 29 (2000).
\bibitem{ppg_CMD}R.~R.~Akhmetshin {\it et al.,} (CMD-2 Collaboration),
Phys.\ Lett.\  B {\bf 562}, 173  (2003).
\bibitem{ppg_SND_phi} 
V.~M.~Aulchenko {\it et al.} (SND Collaboration), 
J.\ Exp.\ Theor.\ Phys.\  {\bf 90}, 927 (2000)
[Zh.\ Eksp.\ Teor.\ Fiz.\  {\bf 117}, 1067 (2000)].
\bibitem{KLOE} 
F.~Ambrosino {\it et al.}  (KLOE Collaboration),
Phys.\ Lett.\ B {\bf 669}, 223 (2008).
\bibitem{SND2010}
M.~N.~Achasov {\it et al.} (SND Collaboration),
JETP Lett.\  {\bf 94}, 2 (2012).
\bibitem{dm2}
D.~Bisello {\it et al.,}
Nucl.\ Phys.\ Proc.\ Suppl.\  {\bf 21}, 111 (1991).
\bibitem{4pi_snd}M.~N.~Achasov {\it et al.} (SND Collaboration),
JETP {\bf 96}, 789 (2003).
\bibitem{4pi_cmd}R.~R.~Akhmetshin {\it et al.} (CMD-2 Collaboration)
Phys.\ Lett.\ B {\bf 466}, 392 (1999).

\bibitem{compton}E.~V.~Abakumova {\it et al.,}
Phys.\ Rev.\ Lett.\  {\bf 110}, 140402 (2013)
arXiv:1211.0103 [physics.acc-ph].

\bibitem{Berends}
F.~A.~Berends and R.~Kleiss,
Nucl.\ Phys.\ B {\bf 186}, 22 (1981).
\bibitem{pdg}
J. Beringer {\it et al.} (Particle Data Group),
Phys. Rev. D {\bf 86}, 010001 (2012).
\bibitem{KEYS} K.~S.~Cranmer,
Comput.\ Phys.\ Commun.\  {\bf 136}, 198 (2001).


\bibitem{5pi}
B.~Aubert {\it et al.}  (BABAR Collaboration),
Phys.\ Rev.\ D {\bf 76}, 092005 (2007)
[Erratum-ibid.\ D {\bf 77}, 119902 (2008)].
\bibitem{babar_3pi} 
B.~Aubert {\it et al.}  (BABAR Collaboration),
Phys.\ Rev.\ D {\bf 70}, 072004 (2004).

\bibitem{FadinRad} E.~A.~Kuraev and V.~S.~Fadin,
Sov.\ J.\ Nucl.\ Phys.\  {\bf 41}, 466 (1985)
[Yad.\ Fiz.\  {\bf 41}, 733 (1985)].
\bibitem{BoneMartine}G.~Bonneau and F.~Martin,
Nucl. Phys. B {\bf 27}, 381 (1971).
\bibitem{lansberg}L.~G.~Landsberg,
Phys.\ Rept.\  {\bf 128}, 301 (1985).
\bibitem{pacetti}S.~Pacetti,
Eur.\ Phys.\ J.\ A {\bf 38}, 331 (2008).
\bibitem{CMDomega}R.~R.~Akhmetshin {\it et al.} (CMD-2 Collaboration),
Phys.\ Lett.\ B {\bf 613}, 29 (2005).
\bibitem{SNDomega}M.~N.~Achasov {\it et al.} (SND Collaboration),
JETP\  {\bf 107}, 61 (2008).
\bibitem{LeptonG}R.~I.~Dzhelyadin {\it et al.}
Phys.\ Lett.\ B {\bf 102}, 296 (1981)
[JETP Lett.\  {\bf 33}, 228 (1981)].
\bibitem{NA60}R.~Arnaldi {\it et al.} (NA60 Collaboration),
Phys.\ Lett.\ B {\bf 677}, 260 (2009).

\bibitem{cleo}K.~W.~Edwards {\it et al.}  (CLEO Collaboration),
Phys.\ Rev.\ D {\bf 61}, 072003 (2000).

\end{thebibliography}
 \end{document}